\documentclass{ieeeaccess}
\usepackage{silence}
\usepackage{cite}
\usepackage{amsmath,amssymb,amsfonts}
\usepackage{amsthm} 
\usepackage{float}
\usepackage[normalem]{ulem}
\usepackage{nccmath}
\usepackage{graphicx}
\usepackage{setspace}
\usepackage{color}
\usepackage{mathptmx}
\usepackage{tgtermes}
\usepackage{caption}
\usepackage{multirow}
\usepackage{multicol}
\usepackage{adjustbox}
\usepackage{kotex}
\usepackage[utf8]{inputenc} 
\usepackage{kotex} 
\usepackage{soul}
\usepackage{xspace}
\usepackage{tgtermes}
\usepackage{verbatim}
\usepackage{booktabs}
\usepackage{mathptmx}
\usepackage{algorithm} 
\usepackage{subfigure}
\usepackage{algpseudocode}
\usepackage{indentfirst}
\usepackage{booktabs}
\usepackage{kotex}
\usepackage{algorithm}
\usepackage{algpseudocode}
\usepackage{grffile}
\usepackage{multirow}
\usepackage{rotating}
\usepackage{mathtools} 
\usepackage{color}
\usepackage{array}
\usepackage{tabulary}      
\usepackage{float}
\usepackage{rotating} 
\usepackage{stfloats}
\usepackage{url}
\usepackage{soul}
\usepackage{textcomp}

\usepackage{balance}
\usepackage{booktabs, multirow}
\usepackage[algo2e, linesnumbered, ruled, vlined]
{algorithm2e}
\usepackage{etoolbox}
\def\textbf#1{{\bfseries #1}}

\def\BibTeX{{\rm B\kern-.05em{\sc i\kern-.025em b}\kern-.08em
    T\kern-.1667em\lower.7ex\hbox{E}\kern-.125emX}}
\begin{document}
\history{Date of publication xxxx 00, 0000, date of current version xxxx 00, 0000.}
\doi{00.0000/ACCESS.2023.0000000}

\title{Entropy-Aware Similarity for Balanced Clustering: A Case Study with Melanoma Detection}
\author{
    \uppercase{Seok Bin Son}\authorrefmark{1}, 
    \uppercase{Soohyun Park} \authorrefmark{1},
    and
    \uppercase{Joongheon Kim}\authorrefmark{1} \IEEEmembership{Senior Member, IEEE} 
}

\address[1]{Department of Electrical and Computer Engineering, Korea University, Seoul, Republic of Korea (e-mails: lydiasb@korea.ac.kr, joongheon@korea.ac.kr)}
\tfootnote{This work was supported by Institute of Information \& Communications Technology Planning \& Evaluation (IITP) grant funded by the Korea Government (MSIT) (No. 2021-0-00766, Development of Integrated Development Framework that supports Automatic Neural Network Generation and Deployment optimized for Runtime Environment).}

\markboth
{S. B. Son \headeretal: Entropy-Aware Similarity for Balanced Clustering: A Case Study with Melanoma Detection}
{S. B. Son \headeretal: Entropy-Aware Similarity for Balanced Clustering: A Case Study with Melanoma Detection}

\corresp{Corresponding authors: Soohyun Park (e-mail: soohyun828@korea.ac.kr).}

\begin{abstract}
Clustering data is an unsupervised learning approach that aims to divide a set of data points into multiple groups. It is a crucial yet demanding subject in machine learning and data mining. Its successful applications span various fields. However, conventional clustering techniques necessitate the consideration of balance significance in specific applications.
Therefore, this paper addresses the challenge of imbalanced clustering problems and presents a new method for balanced clustering by utilizing entropy-aware similarity, which can be defined as the degree of balances. We have coined the term, entropy-aware similarity for balanced clustering (EASB), which maximizes balance during clustering by complementary clustering of unbalanced data and incorporating entropy in a novel similarity formula that accounts for both angular differences and distances. The effectiveness of the proposed approach is evaluated on actual melanoma medial data, specifically the International Skin Imaging Collaboration (ISIC) 2019 and 2020 challenge datasets, to demonstrate how it can successfully cluster the data while preserving balance. Lastly, we can confirm that the proposed method exhibited outstanding performance in detecting melanoma, comparing to classical methods.
\end{abstract}

\begin{IEEEkeywords}
Balanced Clustering, Cluster, Entropy, Melanoma Detection
\end{IEEEkeywords}

\titlepgskip=-21pt

\maketitle

\section{Introduction}

\subsection{Background and Motivation}

\textbf{Clustering} aims to group comparable data points using similarity measures~\cite{cl_kim2007coverage,cl_chen2017dheat,cl_das2007automatic,cl_he2017pattern,cl_wang2011svstream}. Clustering is a significant and challenging issue in the fields of machine learning and data mining. In recent decades, numerous clustering algorithms have emerged, utilizing diverse methods, such as cosine similarity~\cite{cos_li2019improved}, and Euclidean distance~\cite{ec_wang2005euclidean, singh2019speaker}. Various clustering methods display varied performances on various datasets, each with its own set of strengths and limitations. 

In clustering, achieving balance is a challenging task in many applications. If the clusters are imbalanced to each other, clustering may not display the properties of the underlying data. Furthermore, the data can be vulnerable to overfitting or underfitting. However, it is hard to find appropriate algorithms in terms of balanced clustering for specific data distributions and applications, such as our considering melanoma detection.

Melanoma is a fatal type of skin cancers emerged by malignant tumors resulting from melanin and skin pigment cells~\cite{balch2001final, freedberg1999screening, rigel1996incidence}. Gender and Age are two critical determinants of the incidence and death of melanoma, among other factors. Many prior studies show that melanoma mortality is higher among men than women and older age groups have a higher mortality rate~\cite{nosrati2014sex, sreelatha2019early, olsen2020evaluation, balch2001prognostic, chao2004correlation}. In addition, melanoma patients' survival rates are linked to their time of diagnosis~\cite{voss2015improving}.

\begin{figure*}[t!]
\centering
    \includegraphics[width=\textwidth]{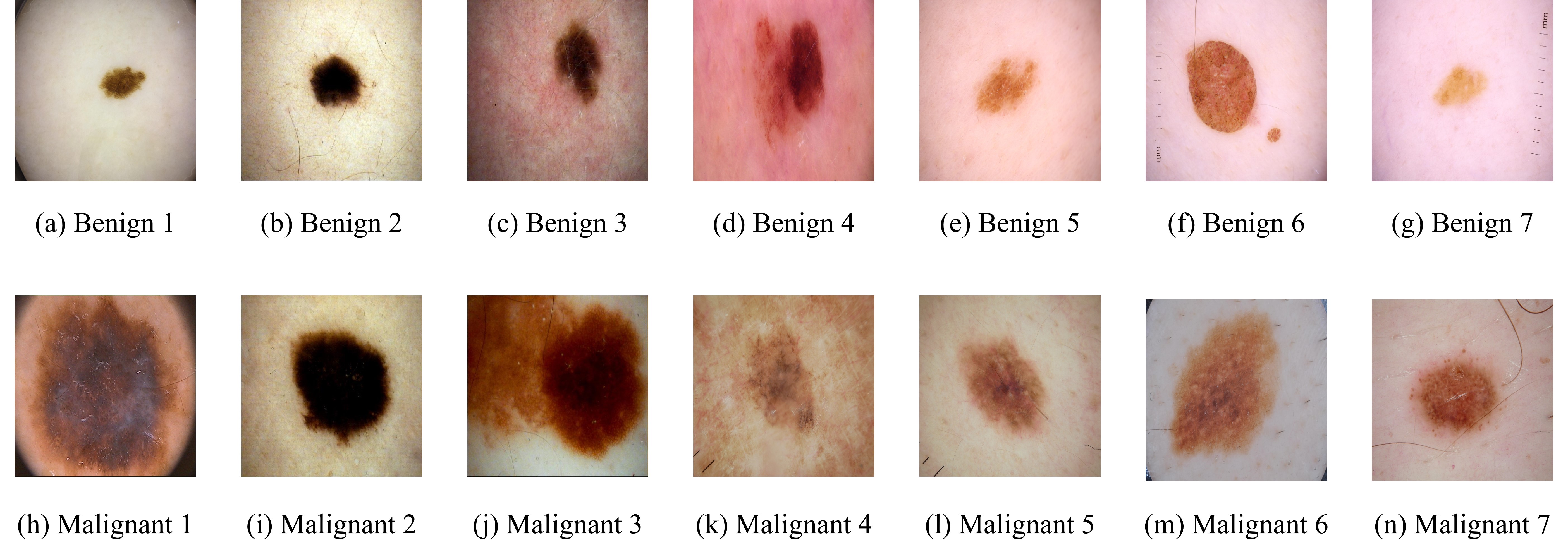}
    \caption{Examples of image datasets of melanoma from the International Skin Imaging Collaboration (ISIC) 2019 and 2020 challenge datasets~\cite{2019_1,2019_2,2019_3,2020}; (a)-(g): benign and (h)-(n): malignant, respectively.}
    \label{malanoma}
\end{figure*}

Early diagnosis of melanoma is crucial for improving survival rates, but accurate diagnosis is challenging due to the complex visual characteristics of skin lesions, such as the shape and size of the disease, as shown in Fig.~\ref{malanoma}~\cite{adegun2019enhanced}. Therefore, in recent years, computer-aided diagnosis (CAD) has been used in various medical fields~\cite{ha2021spatio, jeon2019privacy, ha2022feasibility, yun2022hierarchical}, and moreover, it has also been used to detect melanoma that requires early and accurate diagnosis~\cite{adegun2019deep}.

However, the accurate diagnosis of melanoma through automated computer-based deep learning and machine learning algorithms depends on the availability of unbiased and comprehensive medical datasets. The limitations to gathering data from a single hospital include biased information based on labeling, patient gender, and age, which can lead to inadequately trained methods. Moreover, this situation can be further significant because medical data cannot be shared among the research institutes or hospitals due to patients' privacy, which can lead to data imbalance and data insufficiency. 
For more details, obtaining melanoma datasets from multiple hospitals presents challenges related to personal information privacy, making it nearly impossible to gather all actual patient data. In addition, implementing deep neural networks in hospitals to train algorithms using them is also a challenging task. Furthermore, achieving balance between clusters is also a crucial and fundamentally important task which utilizes melanoma datasets, influenced by important factors such as gender and age.

\subsection{Design Concept}  
This paper proposes a novel inter-hospital clustering method, which is named to \textit{entropy-aware similarity for balanced clustering (EASB)}. Our proposed EASB algorithm focuses on achieving balance between clusters. In addition, the proposed EASB algorithm considers the reality of data collection and privacy aspects to improve the early detection and accurate diagnosis of melanoma. The algorithm assumes that data sharing is possible within the clusters, which protects personal privacy information while providing sufficient data quantity for learning because hospitals do not have to share information with all the other hospitals. The proposed EASB algorithm can also address the issue of biased information based on labeling, patient gender, and age, that issues can arise from gathering data from a single hospital. Moreover, the clustering method can provide the necessary data quantity for hospitals those are with insufficient melanoma data.

\subsection{Contributions}
The main contributions of this research results can be summarized as follows.
\begin{itemize}
    \item This paper proposes a novel clustering algorithm that prioritizes balance within the group while clustering based on the balance of gender and age. The proposed algorithm utilizes entropy to cluster unbalanced data in a complementary manner, resulting in improved balance within the cluster, and in turn, maximized balance degree.
    
    \item The proposed EASB algorithm includes a new similarity measurement that accounts for angular difference, distance, and also balance for more realistic measurements.

    \item In addition, the proposed EASB algorithm can solve the problem of biased information collection due to data collection in a single hospital by the balanced clustering of multiple hospitals.

    \item Moreover, the proposed EASB algorithm preserves patients' privacy for hospital datasets by eliminating the necessity for hospitals to share information with all other hospitals.

    \item Furthermore, the evaluation of our proposed approach on real-world case study for melanoma detection applications is conducted, specifically with hospital-gathered datasets used in ISIC 2019~\cite{2019_1,2019_2,2019_3} and 2020~\cite{2020} challenge. It demonstrates that our algorithm outperforms existing clustering approaches in terms of clustering balance.

    \item Lastly, the proposed EASB algorithm consistently outperforms the most balanced hospital in each scenario in melanoma detection applications. Therefore, this paper offers a new fundamental approach for melanoma detection through clustering that can enhance the both of accuracy as well as effectiveness of melanoma diagnosis.
\end{itemize}

\subsection{Organization} The rest of this paper is organized as follows. Sec.~\ref{sec:related} introduces our related work. Sec.~\ref{sec:Balance} presents our approach, entropy-aware similarity for balanced clustering in melanoma detection applications. In Sec.~\ref{sec:performance}, we present the performance evaluation, and finally, Sec.~\ref{sec:conclusion} concludes this paper.

\section{Related work}\label{sec:related}

\subsection{Similarity-based clustering methods}
The goal of clustering is to categorize similar data points using conventional clustering algorithms and similarity measures~\cite{cl_wang2015multi,cl_wu2017euler,cl_yang2014multitask,cl_zhang2017fast,cl_frey2007clustering,cl_wang2013multi}. The traditional similarity-based clustering algorithms could be classified into angle-based and distance-based similarities. 
The approach that measures similarity based on the cosine angle between two vectors is called angle-based similarity, and it is the most widely used algorithm. As a result, the similarity is assessed by whether the two vectors approximately point in the same direction depending on the angle between the two vectors. The two vectors are increasingly similar and closer to being identical to the value 1 when the angle between them is reduced. The similarity between the two vectors decreases with increasing angle; at this point, it is close to -1. This conclusion only depends on the direction of the vector and is independent to the scale/magnitude of vectors and the distances between the vectors~\cite{cos_li2019improved}. This can be formulated as follows,
\begin{align}
    d_{c}(A, B) &= \frac{A\cdot B}{\left \|A\left \|  \right \|B\right \|}  \\ &= \frac{\sum_{i=1}^{n}A_i \times B_i}{\sqrt{\sum_{i=1}^{n} (A_i)^2} \times \sqrt{\sum_{i=1}^{n} (B_i)^2}} 
        \label{eq:cosine}
\end{align}

where $d_{c}(A, B)$, \textbf{$A$}, \textbf{$B$}, $\|\cdot\|$, \textbf{$A_i$}, \textbf{$B_i$}, and $n$ stand for the cosine similarity between two vectors \textbf{$A$} and \textbf{$B$}, vector \textbf{$A$}, vector \textbf{$B$}, the magnitude, $i$-th element of vector \textbf{$A$}, $i$-th element of vector \textbf{$B$}, and the size of vectors.

On the other hand, the distance-based similarity is a commonly used function in machine learning, data mining, and information retrieval domains. Specifically, Euclidean distance is frequently employed in clustering to comparable group datasets. Euclidean distance is calculated as the straight-line distance between two points in Euclidean space, derived from the square root of the summation of the squared differences in each dimension~\cite{ec_wang2005euclidean, singh2019speaker}. Regarding similarity measurement, the Euclidean distance between two data points can be used to determine their similarity, where a smaller Euclidean distance indicates greater proximity between the points. The corresponding computation procedure can be denoted as,
\begin{equation}
    \label{eq:eu}
    d_{\textit{e}}(A, B) = \sqrt{\sum_{i=1}^{n}(A_i - B_i)^2} 
\end{equation}
where $d_{\textit{e}}(A, B)$, \textbf{$A_i$}, \textbf{$B_i$}, and $n$ stand for the Euclidean distance between two vectors \textbf{$A$} and \textbf{$B$}, $i$-th element of vector \textbf{$A$}, $i$-th element of vector \textbf{$B$}, and the size of vectors.

\subsection{Melanoma detection}

Melanoma is a kind of skin cancer caused by sun exposure that has a low survival rate of 15-20\%~\cite{albahli2020melanoma}. Early detection is essential for improving survival rates in cases of melanoma, a skin condition with a high fatality rate. Melanoma can impact different organs, making diagnosis much more difficult if it is not discovered early. Melanoma lesions can be difficult to detect because of various issues, including different skin tones, hazy borders, intricate backgrounds, and different structures and regions~\cite{sreelatha2019early}. Although numerous factors influence melanoma incidence and death, gender and age are essential. Men are more likely than women to get melanoma globally, and men tend to die from the disease at much higher rates than women~\cite{nosrati2014sex, sreelatha2019early, olsen2020evaluation}. Melanoma is the second leading cause of death in adults over 65 in the United States, and studies have indicated that patients with the disease have a poor prognosis after 60 years of age~\cite{balch2000long, yancik2000aging}.

Many AI techniques have been applied to accurately diagnose melanoma, including the Lesion-classifier technique, which uses an improved encoder-decoder network to detect melanoma pixel by pixel~\cite{adegun2019deep}. In addition, it has been noted that the fuzzy-based GrabCut-stacked convolutional neural networks (GC-SCNN) model performs better for melanoma identification in terms of accuracy and speed~\cite{bhimavarapu2022skin}. Moreover, it has been reported that using a Siamese structure based deep network with a segmentation loss as a regularization term achieve promising results on detecting automatically short-term lesion changes in melanoma screening~\cite{zhang2020short}. Furthermore, utilizing the YOLOv4 object detector to discriminate the infected and non-infected regions have reported that it achieves high accuracy rates and demonstrating the practical applicability of the method in developing clinical decision support systems for melanoma diagnosis~\cite{albahli2020melanoma}. These techniques can help with early detection, improving the chances of survival for melanoma patients.

\begin{figure*}[t!]
\centering
    \includegraphics[width=0.9\textwidth]{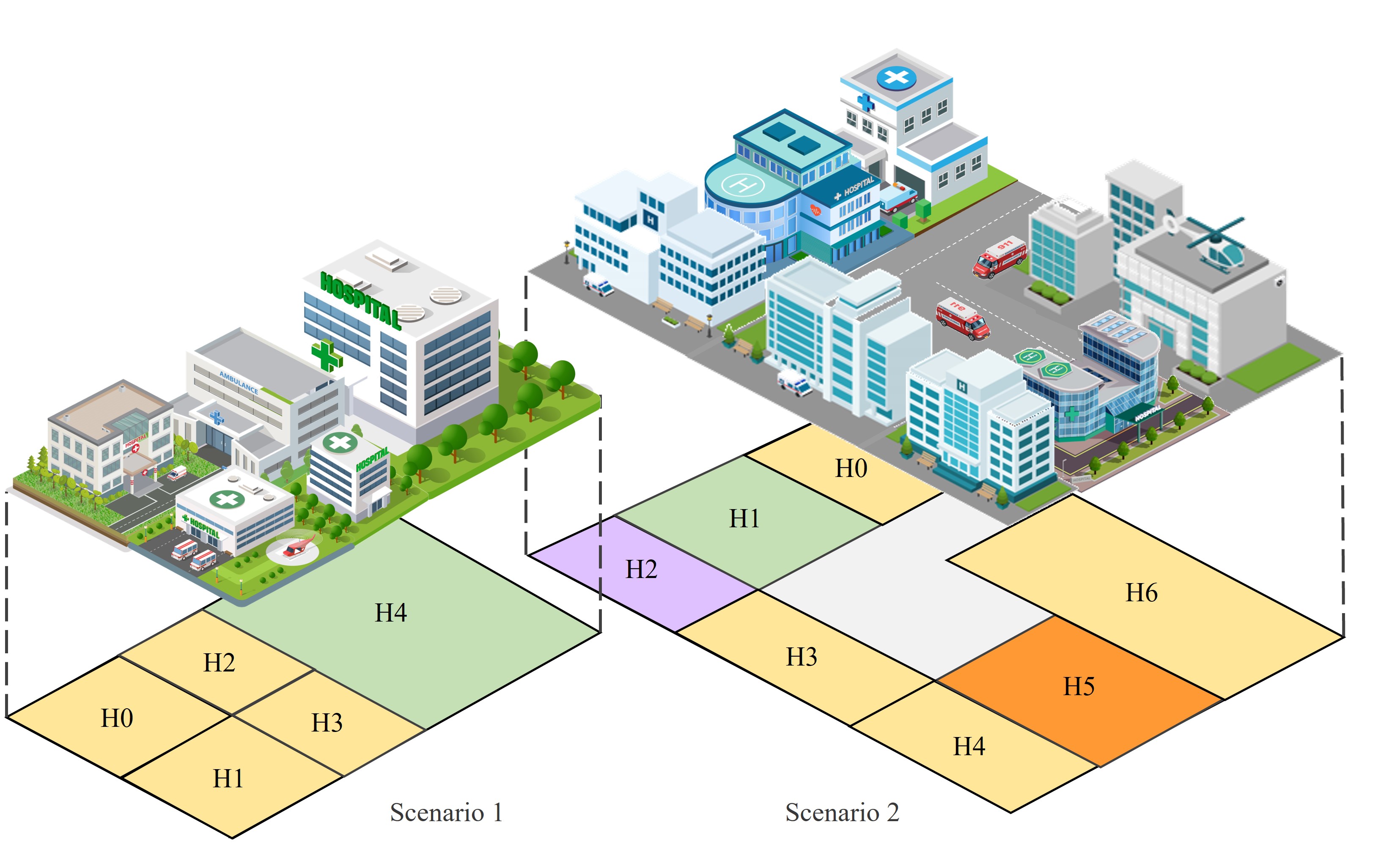}
    \caption{Overall architecture: Scenarios 1 and 2 consist of 5 and 7 hospitals, respectively. According to our method, hospitals are clustered by colors in the figure.}
    \label{fig: overview}
\end{figure*}

\begin{figure*}[t!]
\centering
\includegraphics[width=0.8\textwidth]{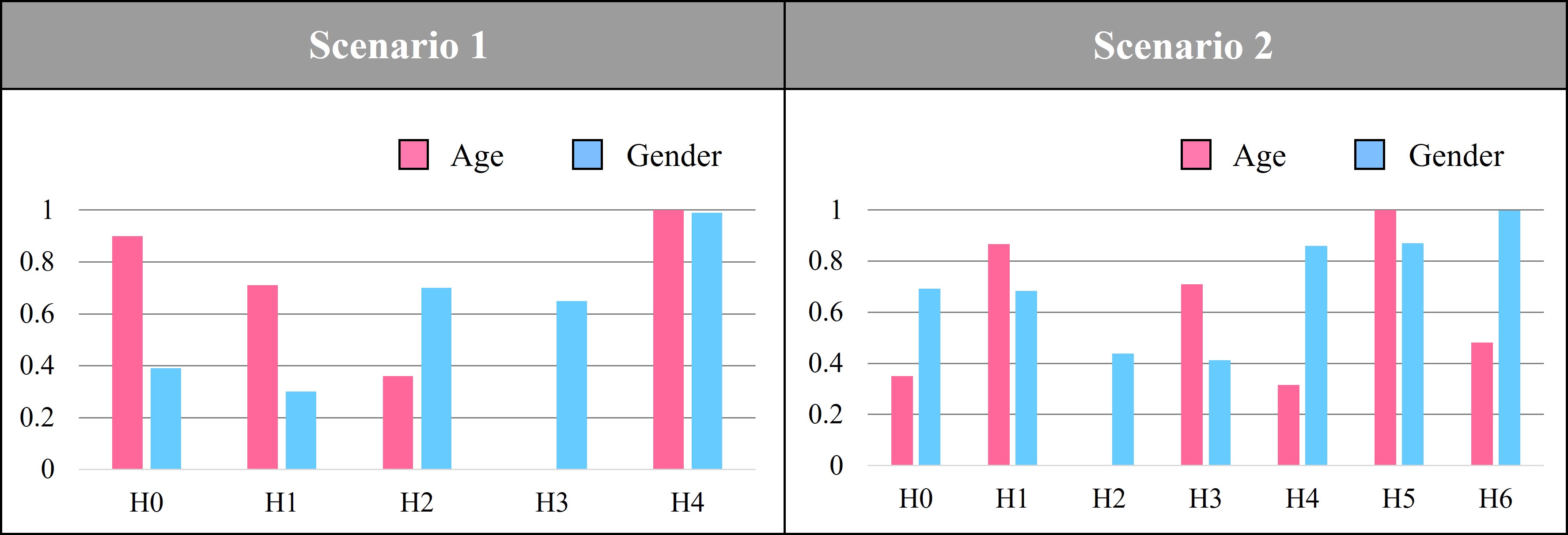}
\caption{Degree of balance: The degree of balance by gender and age of hospital in each scenario.}
\label{degree}
\end{figure*}

\begin{table}[t!]
\centering
\caption{Hospital Distribution in Scenario 1}
\label{tab:Hospital Distribution Detail in Scenario 1}
\begin{small}
\resizebox{0.8\columnwidth}{!}{%
\begin{tabular}{c|c|c}
\hline
\textbf{Hospital}& \textbf{$H_{Age}[Y]_i$ (\%)} & \textbf{$H_{Gender}[Y]_i$ (\%)} \\ \hline \hline
H0 & 90.36                 & 38.52                    \\ \hline
H1 & 71.12                 & 29.74                    \\ \hline
H2 & 36.06                 & 69.39                    \\ \hline
H3 & 0                    & 65.20                    \\ \hline
H4 & 1                    & 99.86                    \\ \hline
\end{tabular}%

}
\end{small}
\end{table}

\begin{table}[t!]
\centering
\caption{Hospital Distribution in Scenario 2}
\label{tab:Hospital Distribution Detail in Scenario 2}
\resizebox{0.8\columnwidth}{!}{%
\small{
\begin{tabular}{c|c|c}
\hline
\textbf{Hospital}& \textbf{$H_{Age}[Y]_i$ (\%)} & \textbf{$H_{Gender}[Y]_i$ (\%)} \\ \hline \hline
H0 &35.11                  &69.22                     \\ \hline
H1 &86.61                  &68.37                     \\ \hline
H2 &0                  &43.87                     \\ \hline
H3 &70.98                    &41.26                     \\ \hline
H4 &31.65                    &85.91                     \\ \hline
H5 &100                     & 87.00                    \\ \hline
H6 &48.17                     &99.86                     \\ \hline
\end{tabular}%
}}
\end{table}

\section{Entropy-Aware Similarity for Balanced Clustering} \label{sec:Balance}

\subsection{Design Motivation}
Preserving balance while clustering data is critical for ensuring each cluster contains comparable data points, accurately representing the underlying data distribution. Unbalanced clusters may result in clustering that does not accurately reflect the actual data distribution, increasing the possibility of overfitting or underfitting during prediction or classification. This paper presents an entropy-based data balance similarity algorithm to solve data imbalance between clusters.

\begin{figure*}[!t]
\centering
\subfigure[]{
\includegraphics[width=.9\columnwidth]{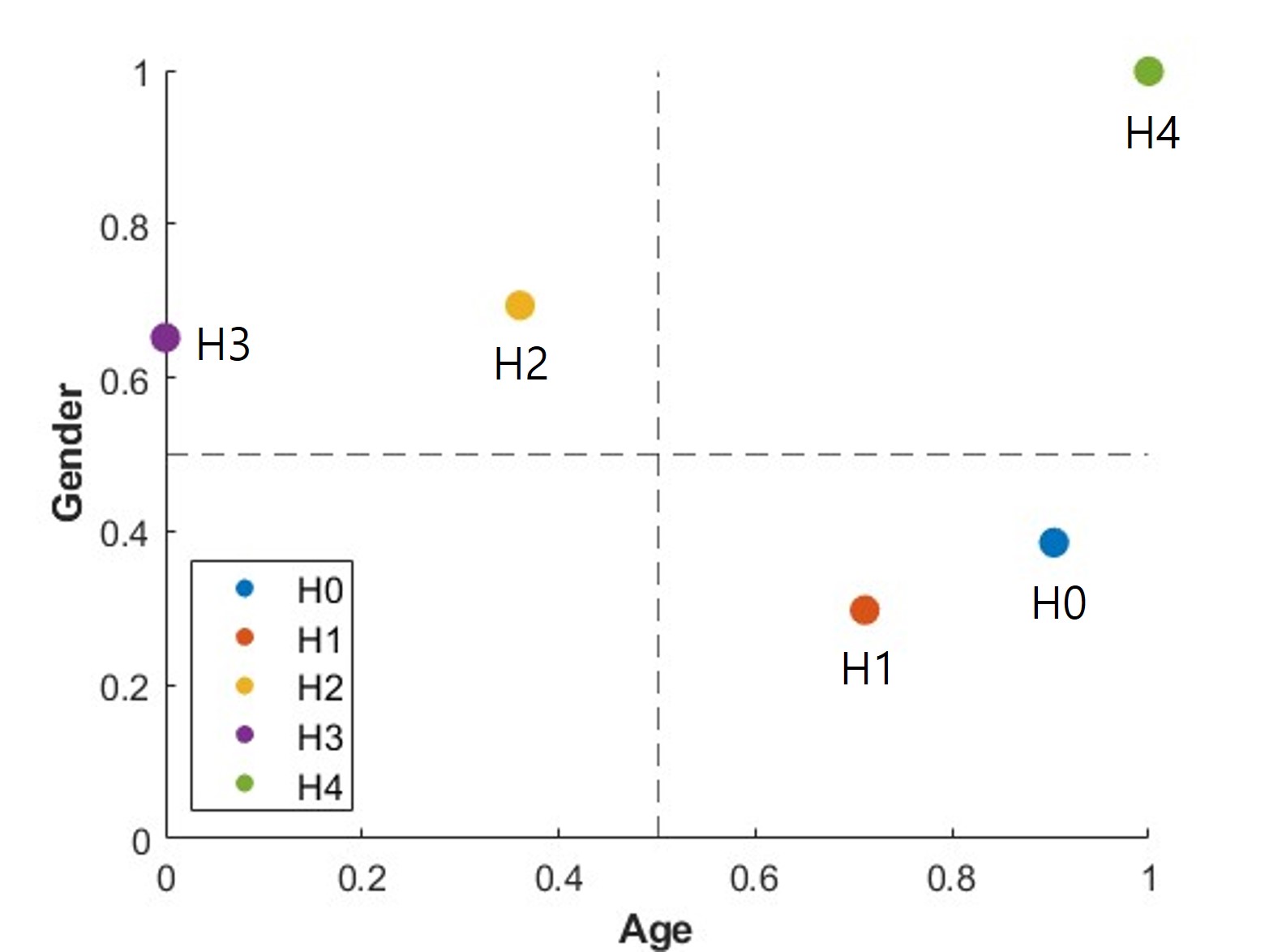}
\label{fig:fig4a}
}
\subfigure[]{
\includegraphics[width=.9\columnwidth]{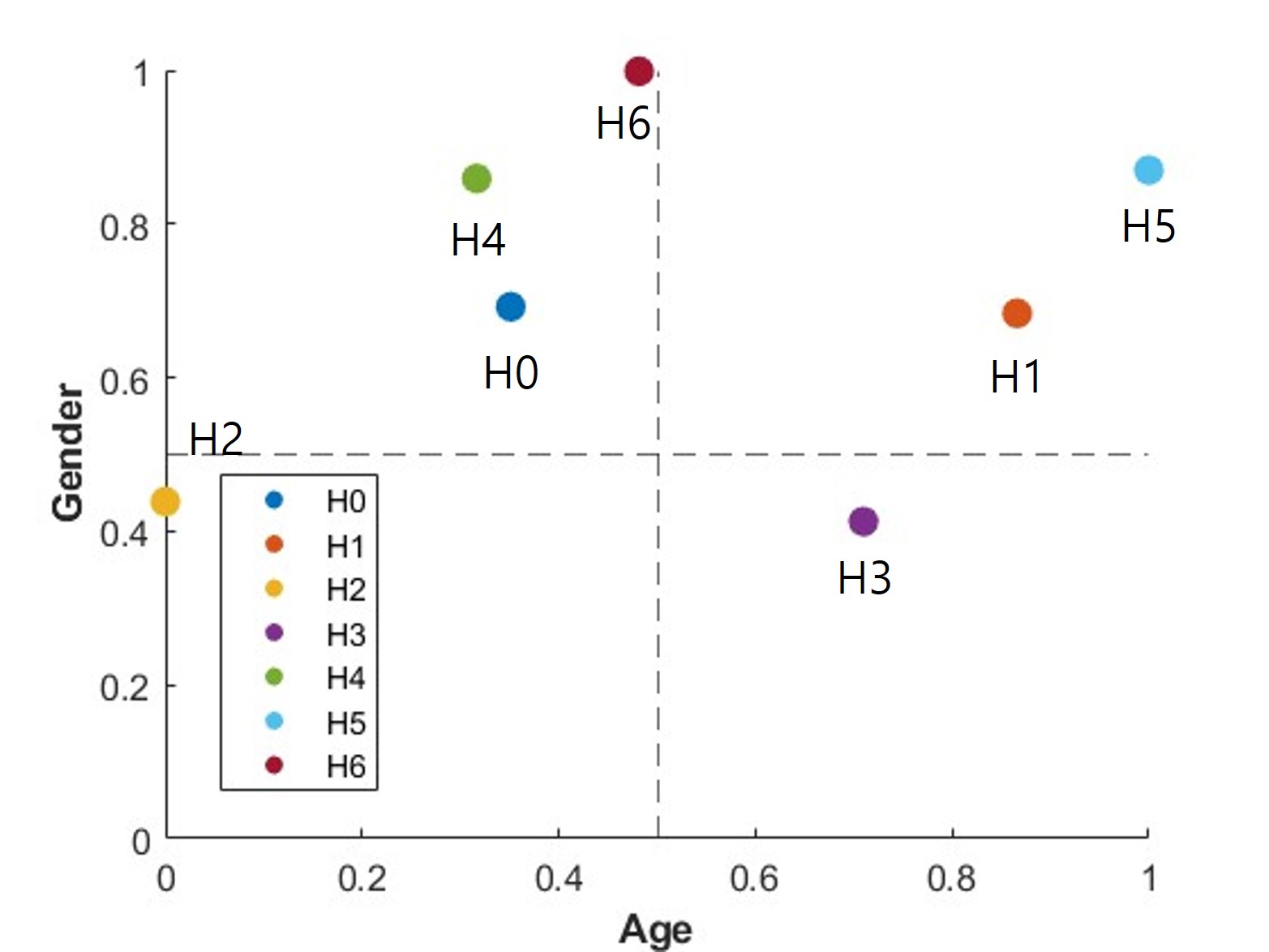}
\label{fig:fig4b}
}
\caption{
Distribution of Hospitals: (a) Scenario 1 and (b) Scenario 2.
}
\end{figure*}

\subsection{Basic Assumption}
This section illustrates our basic assumptions. Our balance similarity algorithm aims to cluster hospitals to address the imbalances of gender and age, significantly impacting melanoma mortality rates. We demonstrate the efficacy of our proposed algorithm in preserving balance within clusters using two scenarios, as shown in Fig.~\ref{fig: overview}. In addition, 
the varied degrees of gender and age imbalance in that scenarios' hospitals are depicted in Fig.~\ref{degree}. In scenario 1, each hospital's entropy distribution is presented in 
Fig.~\ref{fig:fig4a} and Table~\ref{tab:Hospital Distribution Detail in Scenario 1}. The scenario 1 has five hospitals, where H4 is the most balanced hospital, H2 and H3 have disproportionate ages but balanced gender, and lastly, H0 and H1 have disproportionate gender but balanced ages. Moreover, in scenario 2, the entropy distribution of each hospital is illustrated in Fig.~\ref{fig:fig4b} and Table~\ref{tab:Hospital Distribution Detail in Scenario 2}. The scenario 2 has seven hospitals, where H5 is the most balanced hospital, H1 is moderately balanced. In addition, H0, H4, and H6 have disproportionate age but balanced gender, furthermore, H3 has disproportionate gender but balanced age, and lastly, H2 is imbalanced in both gender and age.

\subsection{Our Proposed Entropy-Aware Similarity for Balance Clustering (ESAB) Algorithm} \label{balance_clustering}

This section describes our proposed entropy-aware similarity for balanced (EASB) clustering algorithm aimed at clustering hospitals to address the gender and age imbalances that significantly impact melanoma mortality rates. 

In order to clarify our proposed EASB clustering algorithm, it should be highlighted that our proposed algorithm consists of two sub-problems to answer following two fundamental questions. 

\begin{itemize}
    \item (1) How to detect the degree of gender an age imbalances in hospitals?
    \item (2) How to develop a novel clustering similarity algorithm that considers the degree of balance?
\end{itemize}

To detect gender and age imbalances, the algorithm utilizes the concept of entropy, which can numerically illustrates uncertainty in datasets, to represent the differences in probability distributions. The concept of entropy for gender and age are applied to hospitals from $1$ to $n$, denoted as $\{H_1, H_2, \cdots, H_n\}$. The detailed mathematical representations are as follows, 
\begin{align}
    \label{eq:gender}
    H_{Gender}[Y]_i &= -\sum_{k=1}^{K} p(y_k)log_2p(y_k) \\
    \label{eq:age}
    H_{Age}[Y]_i &= -\sum_{k=1}^{K} p(y_k)log_2p(y_k)
\end{align}
where $H_{Gender}[Y]_i$, $H_{Age}[Y]_i$, $p(y_k)log_2p(y_k)$, and $K$ stand for the $i$-th hospital entropy of gender, the $i$-th hospital entropy of age, a probability mass function, and the size of classes. The entropy of gender and age, $H_{Gender}[Y]_i$ and $H_{Age}[Y]_i$, are plotted on the $x$-axis and $y$-axis of~\ref{fig:fig4a} and~\ref{fig:fig4b}, respectively. 

We set a weight value indicating the degree of balance in each hospital. The previously defined values of $H_{Gender}[Y]_i$ and $H_{Age}[Y]_i$ are used to calculate the weight values of $HW_i$, with a value closer to 0 indicating a more severe imbalance in the hospital, which can be formulated as, 

\begin{equation}
    \label{eq:HW_i}
    HW_i = H_{Gender}[Y]_i \times H_{Age}[Y]_i
\end{equation}
where $HW_i$, $H_{Gender}[Y]_i$, and $H_{Age}[Y]_i$ stand for the weight value indicating the degree of balance $i$-th hospital, $i$-th hospital entropy of gender, and $i$-th hospital entropy of age, respectively.

In addition, the degree of balance when clustering between two $i$-th and $j$-th hospitals is denoted as $HW_{i,j}$, where this can be mathematically formulated as follows,
\begin{equation}
    \label{exp}
    HW_{i,j} = \frac{1}{1+exp(-z_{i,j})}
\end{equation}
where $z_{i,j}$ can be obtained by taking logarithmic function from $\alpha_{i,j}$, i.e.,
\begin{equation}
    \label{eq:z}
    z_{i,j} = log(\alpha_{i,j})
\end{equation}
and the $\alpha_{i,j}$ stands for the odds ratio of $p(H_{i,j})$, i.e., the average of degree of balance between $i$-th and $j$-th hospitals, i.e., 
\begin{equation}
    \label{eq:odds ratio} 
    \alpha_{i,j} = \frac{p(H_{i,j})}{1-p(H_{i,j})}
\end{equation}
where
\begin{equation}
    \label{eq:p}
    p(H_{i,j}) = \frac{HW_i + HW_j}{2},
\end{equation}
where $HW_{i}$ and $HW_{j}$ can be obtained by \eqref{eq:HW_i}.

Note that, to normalize the corresponding value, a logistic function was used to express it as a value between 0 and 1, as shown in \eqref{eq:z}.

\begin{figure}[h!]
\centering
\subfigure[]{
\includegraphics[width=0.97\columnwidth]{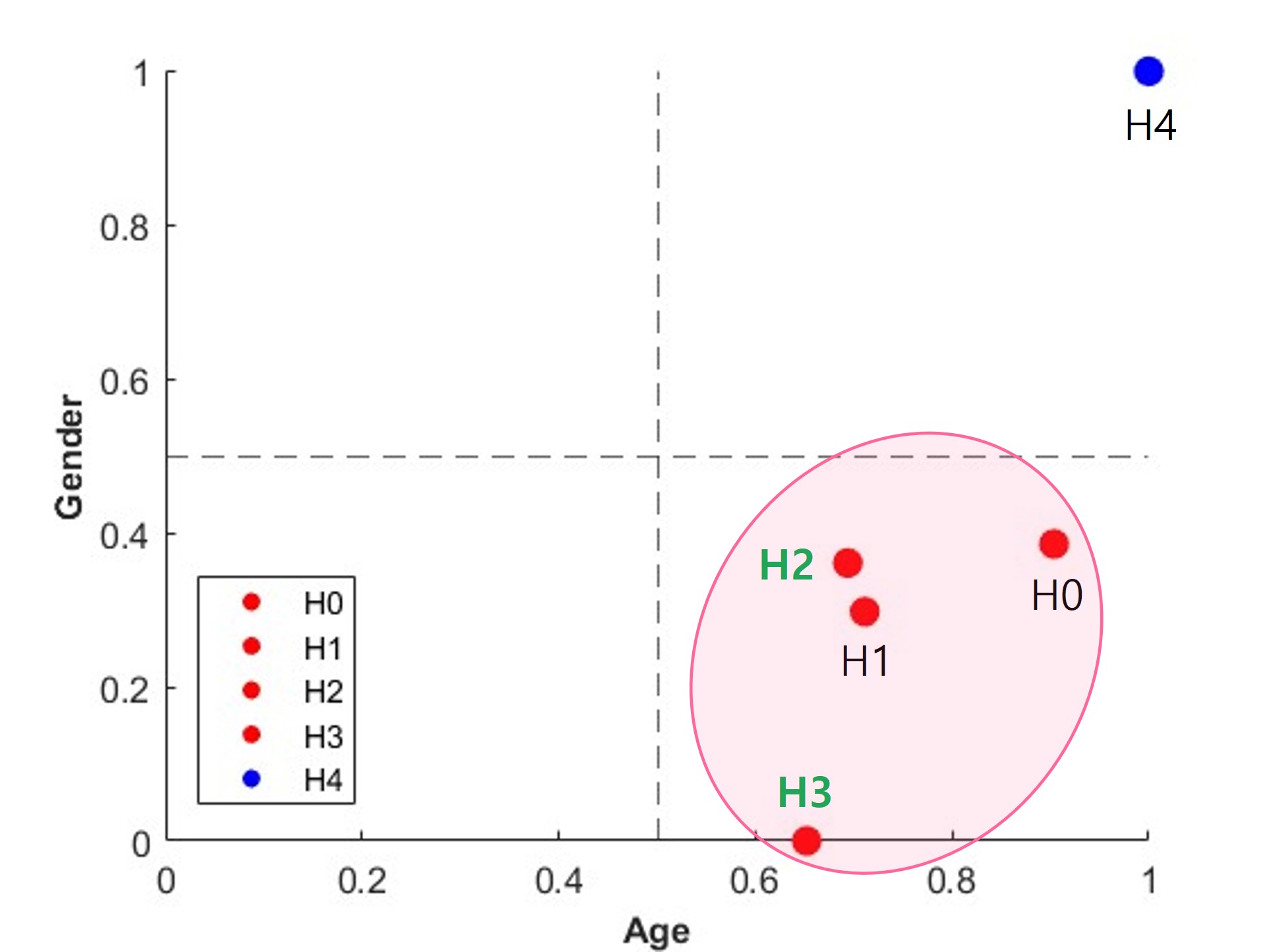}
}
\vspace{0.1cm}
\subfigure[]{
\includegraphics[width=0.97\columnwidth]{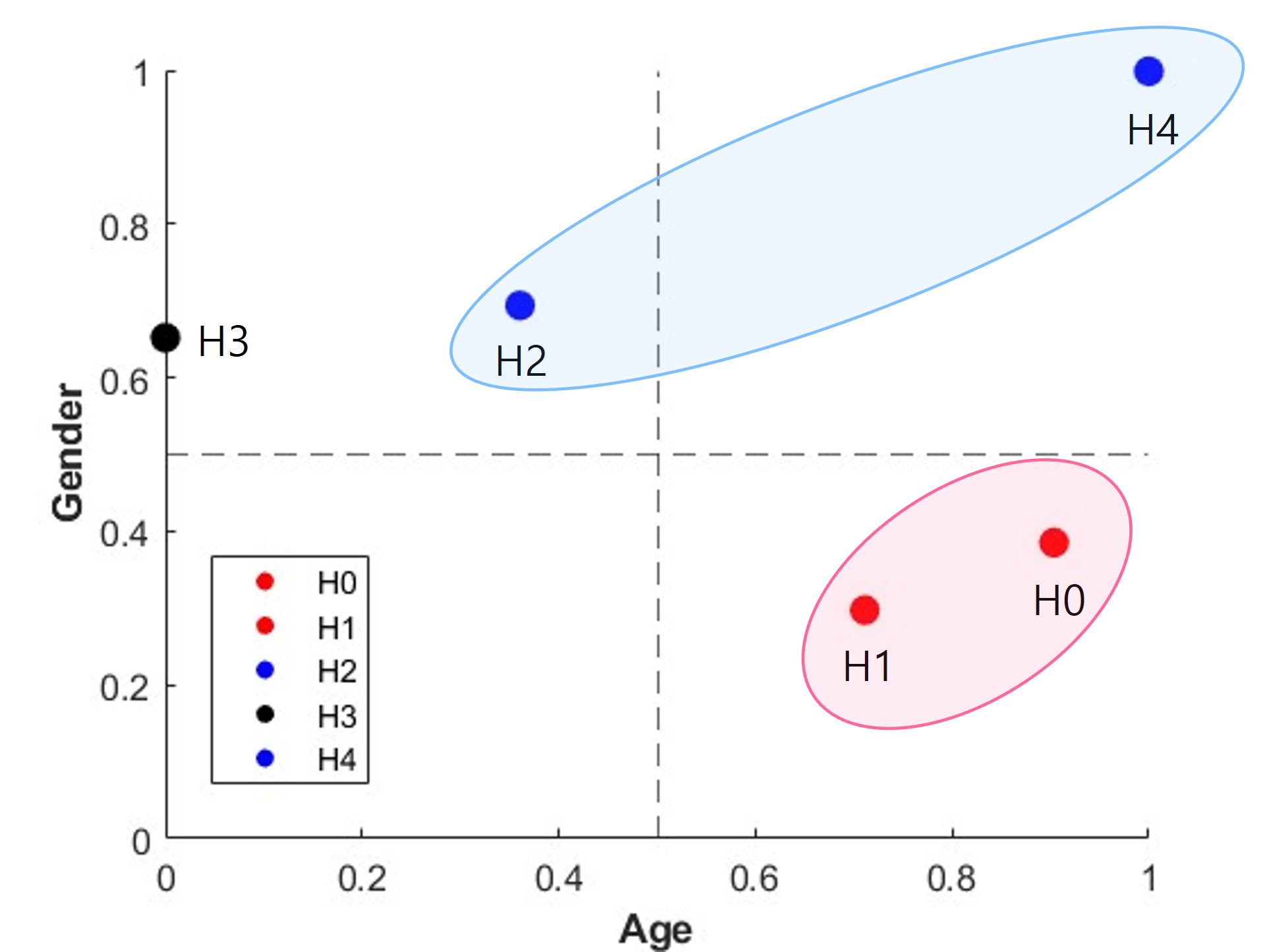}
}
\vspace{0.1cm}
\subfigure[]{
\includegraphics[width=0.97\columnwidth]{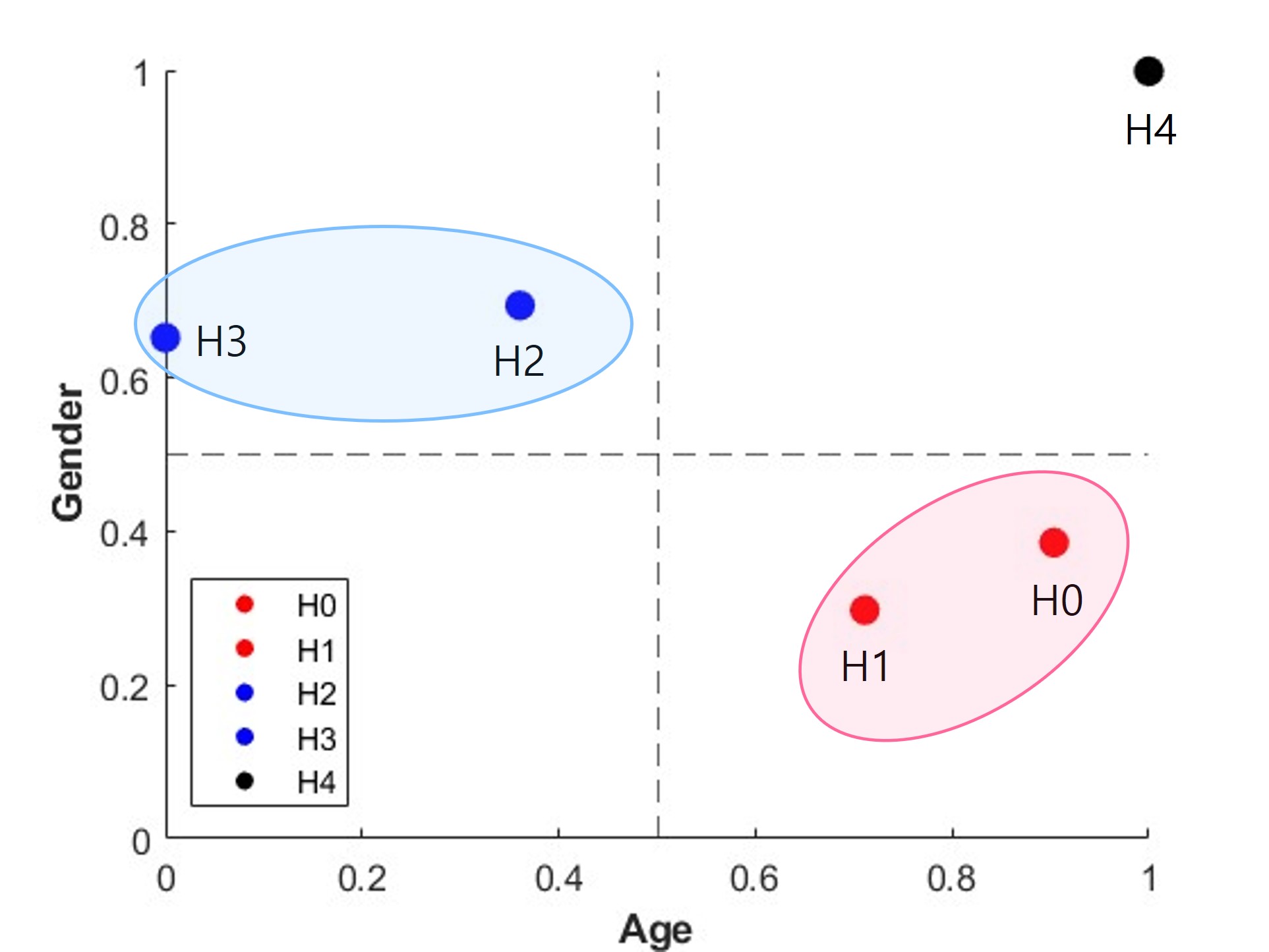}
}
\caption{
Clustering Results in Scenario 1: (a) EASB Clustering Algorithm, (b) Cosine Similarity, and (c) Euclidean Distance.
}
\label{fig:S1}
\end{figure}

We propose a novel clustering similarity that considers balance by transforming the $x$-axis values of hospitals symmetrically into the $y$-axis values of hospitals. The proposed approach is designed to balance groups that are disproportionate in age efficiently but balanced in gender and vice versa in one cluster. By symmetrical transformation, hospitals with opposite properties gather at relatively short distances and angles, increasing the possibility of being grouped into the same cluster when determining similarity based on angle and distance. We illustrates the transformed hospitals in green, as depicted in Fig.~\ref{fig:S1}(a) and Fig.~\ref{fig:S2}(a). The proposed approach hybridizes existing angle-based and distance-based similarity algorithms to generate a more substantial novel similarity for balanced clustering that considers both the similarity of angles and the proximity of data. The inter-hospital balance score is also used to create a new similarity metric, denoted as $d_{EASB}(H_i, H_j)$. Finally, this can be formulated as, 
\begin{equation}
    \label{eq:bs}
    \max : d_{EASB}(H_i, H_j) \\
    = \frac{d_{c}(H_i, H_j) \times HW_{ij}}{1+d_{e}(H_i, H_j)}
\end{equation}
where 
\begin{align}
d_{c}(H_i, H_j) &=\frac{(\sum_{i=1}^{n}H_i \times H_j)}{\sqrt{\sum_{i=1}^{n}(H_i)^2}\times \sqrt{\sum_{j=1}^{n}(H_j)^2}} \\
d_{e}(H_i, H_j) &=\sqrt{\sum_{i,j=1}^{n}(H_i-H_j)^2}
\end{align}
where $d_{EASB}(H_i, H_j)$, $d_{c}(H_i, H_j)$, $d_{e}(H_i, H_j)$, and $HW_{ij}$ stand for the our EASB clustering algorithm between $i$-th and $j$-th hospitals, the cosine similarity between $i$-th and $j$-th hospitals, the Euclidean distance between $i$-th and $j$-th hospitals, and the degree of balance between $i$-th and $j$-th hospitals, respectively.

\section{Performance Evaluation}\label{sec:performance}
This section presents the data-intensive performance evaluation results of our proposed EASB clustering algorithm by the case study with large-scale real-world melanoma datasets. In addition, it considers scenarios 1 and 2 in order to construct the hospital dataset. Moreover, our proposed EASB algorithm compares with existing clustering algorithms. Furthermore, we also demonstrate the similar or superior performance of the grouped clusters by our algorithm to the most balanced hospital datasets. The detailed experimental performance evaluation results by the case study with melanoma detection for each scenario are as follows.

\begin{table}[t!]
\centering
\label{tab:dataset}
\caption{Melanoma Detection Dataset}
\resizebox{0.8\columnwidth}{!}{%
\small{
\begin{tabular}{c|c|c|c}
\hline
\textbf{Dataset} & \textbf{Benign} & \textbf{Malignant} & \textbf{Total}\\ \hline \hline
Train & 26,114          & 4,036              & 30,150\\ \hline
Test  & 6,428           & 1,070              & 7,498\\ \hline
Total  & 32,542           & 5,106              & 37,648\\ \hline
\end{tabular}%
}
}
\end{table}

\subsection{Performance Evaluation Configurations}
The study describes the proposed EASB algorithm and also performs data-intensive experiments using the ISIC 2019 and 2020 melanoma datasets~\cite{2019_1,2019_2,2019_3,2020}. The detailed configuration of the dataset used in the experiment is as shown in Table 3. Since the existing 2020 malignant melanoma dataset is insufficient, we further use the malignant dataset in 2019. In addition, we use the VGG16 model, which shows excellent performance for image classification~\cite{vggsimonyan2014very}. Moreover, traditional clustering methods such as cosine similarity and Euclidean distance algorithms are used for performance comparison. In addition, two hospital deployment scenarios are constructed for experimental analysis. Lastly, AUROC~\cite{17_narkhede2018understanding} and confusion matrix~\cite{18_lipton2014optimal} (accuracy, precision, recall, and F1-score) are used for performance evaluation metrics in order to quantitatively analyze the performance of melanoma detection. In terms of software implementation, the experiments are conducted with Python 3.8.0 and TensorFlow 2.11.0, using two GPUs (NVIDIA Corporation GV100GL), Adam optimizer, learning rate with $10^{-55}$, batch size with $128$, and epoch with $20$.

\subsection{Scenario 1}
\subsubsection{Comparison of Balance}

In scenario 1, the clustering results of two existing methods, cosine similarity and Euclidean distance similarity, in the scenario are presented in Fig.~\ref{fig:S1}(b) and Fig.~\ref{fig:S1}(c), respectively. These methods perform clustering based on angle and distance, respectively, however they do not ensure balanced clustering by crossing unbalanced hospitals. In contrast, our proposed algorithm groups all hospitals except for the most balanced hospital (i.e., H4) into one clustering, resulting in two balanced clusters, as shown in Fig.~\ref{fig:S1}(a). 

Table~\ref{tab:SC1_Comparison} provides each method's average degree of balance. As shown in the Table~\ref{tab:SC1_Comparison}, our proposed algorithm for creating clusters demonstrates superior balance compared to other existing methods, indicating that our algorithm results in better balance when forming clusters. Therefore, our algorithm outperforms in terms of balanced clustering than the existing comparing methods.

\begin{table}[t!]
\centering
\caption{Comparison of Balance in Scenario 1}
\small{
\label{tab:SC1_Comparison}
\resizebox{0.9\columnwidth}{!}{%
\begin{tabular}{c|c|c}
\hline
\textbf{Avg\_Cluster} & \textbf{$H_{Age}[Y]_i$ (\%)} & \textbf{$H_{Gender}[Y]_i$ (\%)} \\ \hline \hline
Cosine             & 59.43 & 65.98 \\ \hline
Euclidean distance & 75.41 & 99.86 \\ \hline
EASB        & 86.20  & 99.86 \\ \hline
\end{tabular}%
}}
\end{table}

\subsubsection{Comparison of Melanoma Detection}

\begin{table}[t!]
\centering
\caption{Comparison of Melanoma Detection in Scenario 1}
\label{tab:Scenario 1_result}
\resizebox{\columnwidth}{!}{%
\begin{tabular}{c|c|cccc}
\hline
\multirow{2}{*}{\textbf{\begin{tabular}[c]{@{}c@{}}\rule{0in}{3.2ex} Hospital\\Cluster \end{tabular}}} &
  \multirow{2}{*}{\textbf{\begin{tabular}[c]{@{}c@{}}\rule{0in}{3.2ex} AUROC\\ \normalfont(\%)\end{tabular}}} &
  \multicolumn{4}{c}{\textbf{Confusion Matrix}} \\ \cline{3-6} 
 &
   &
  \multicolumn{1}{c|}{\textbf{\begin{tabular}[c]{@{}c@{}}Accuracy\\ (\%)\end{tabular}}} &
  \multicolumn{1}{c|}{\textbf{\begin{tabular}[c]{@{}c@{}}Precision\\ (\%)\end{tabular}}} &
  \multicolumn{1}{c|}{\textbf{\begin{tabular}[c]{@{}c@{}}Recall\\ (\%)\end{tabular}}} &
  \textbf{\begin{tabular}[c]{@{}c@{}}F1-score\\ (\%)\end{tabular}} \\ \hline \hline
H4 &
  96.25 &
  \multicolumn{1}{c|}{94.21} & 
  \multicolumn{1}{c|}{92.18} &
  \multicolumn{1}{c|}{64.95} &
  76.21 \\ \hline
EASB &
  97.26 &
  \multicolumn{1}{c|}{95.25} &
  \multicolumn{1}{c|}{92.30} &
  \multicolumn{1}{c|}{72.80} &
  81.40
   \\ \hline
\end{tabular}%
}
\end{table}

Our study aims to evaluate the effectiveness of our algorithm for detecting melanoma by comparing its performance on a well-balanced hospital cluster, which contain the hospitals H0, H1, H2, and H3, and the most balanced hospital cluster, which is denoted as H4. If the cluster we created has similar or superior performance to H4, the most balanced hospital, the cluster would have been well balanced. The experiment results are shown in Table~\ref{tab:Scenario 1_result}. As shown in the Table~\ref{tab:Scenario 1_result}, we analyze the results and find that our cluster outperforms H4. Therefore, we can confirm that the balance was well matched with the new similarity formula we created and that melanoma detection is also well performed. In the case of AUROC, the cluster was 97.26\%, which was superior to H4, i.e., 96.25\%. In the case of the confusion matrix, it is confirmed that the cluster performed better than H4. Specifically, in the case of accuracy, the cluster is 95.25\%, which shows better performance than 94.21\%, that is for H4. In the case of precision, it can be seen that the cluster is 92.30\% whereas H4 is 92.18\%, showing very similar performance. In the case of recall, the cluster was 72.80\%, which is superior to 64.95\%, which is for H4. Finally, in F1-score, it can be seen that the cluster is 81.40\%, which is superior to the performance of H4, i.e., 76.21\%.

Moreover, our research analyzes the performance of H4 and our newly constructed balanced cluster algorithm in detecting melanoma by examining the confusion matrix, as depicted in Fig.~\ref{fig:c1m}(a)/(b). Fig.~\ref{fig:c1m}(a) illustrates that H4 accurately identifies 695 malignant out of 1,070 malignant images however misjudged 375 images as benign. It also correctly identifies 6,369 benign images out of 7,498 however misjudged 59 images as malignant. Fig.~\ref{fig:c1m}(b) shows that our newly constructed cluster algorithm accurately identifies 779 malignant out of 1,070 malignant images, however misjudged 291 as benign. It also correctly identifies 6,363 benign images out of 7,498 however misjudged 65 as malignant. In summary, our results formally and mathematically indicate that our new balanced cluster algorithm outperforms in all possible situations for the case study with melanoma detection.

\begin{figure}[!t]
\centering
\subfigure[]{
\includegraphics[width=0.97\columnwidth]{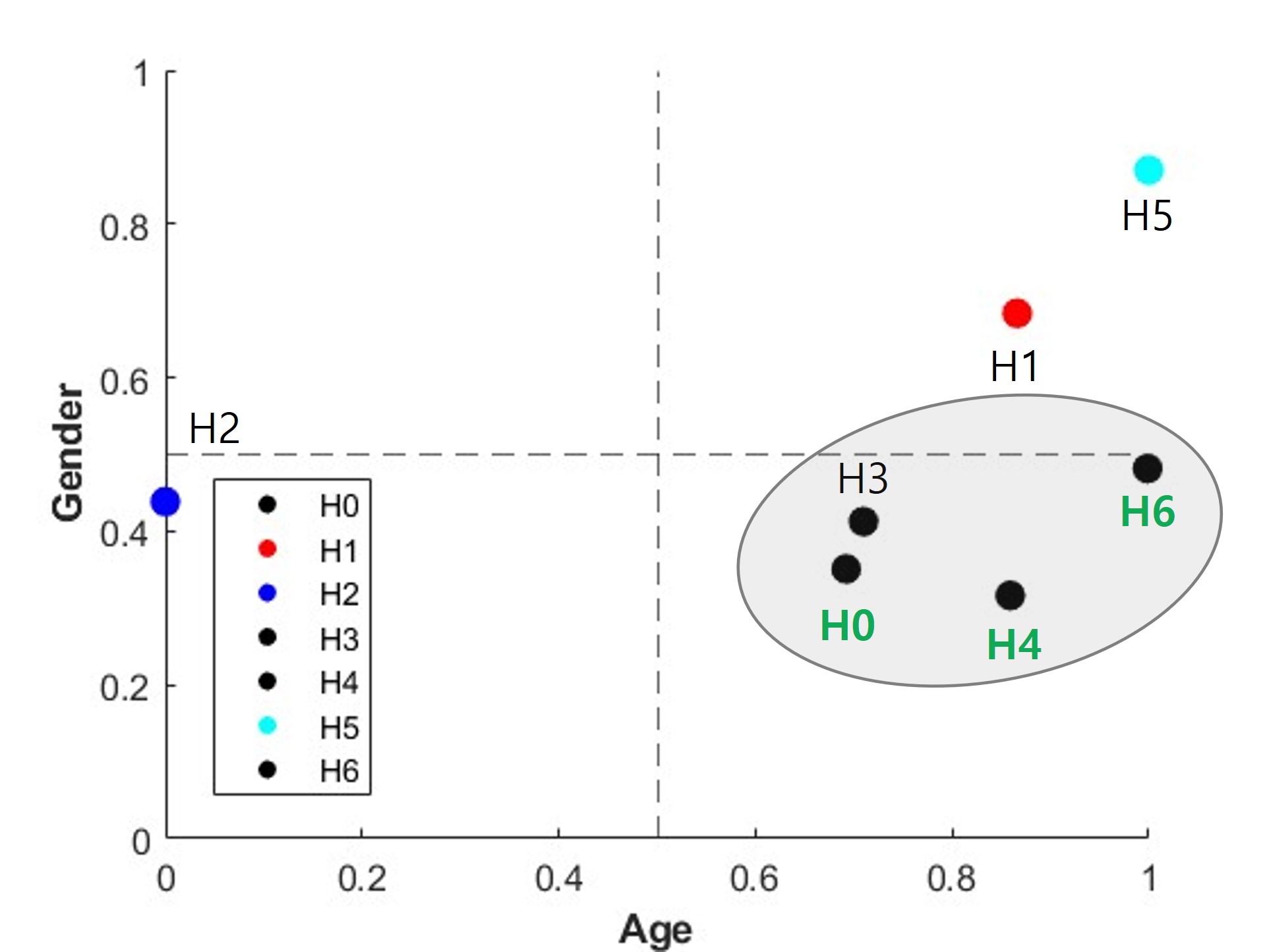}
}
\vspace{0.1cm}
\subfigure[]{
\includegraphics[width=0.97\columnwidth]{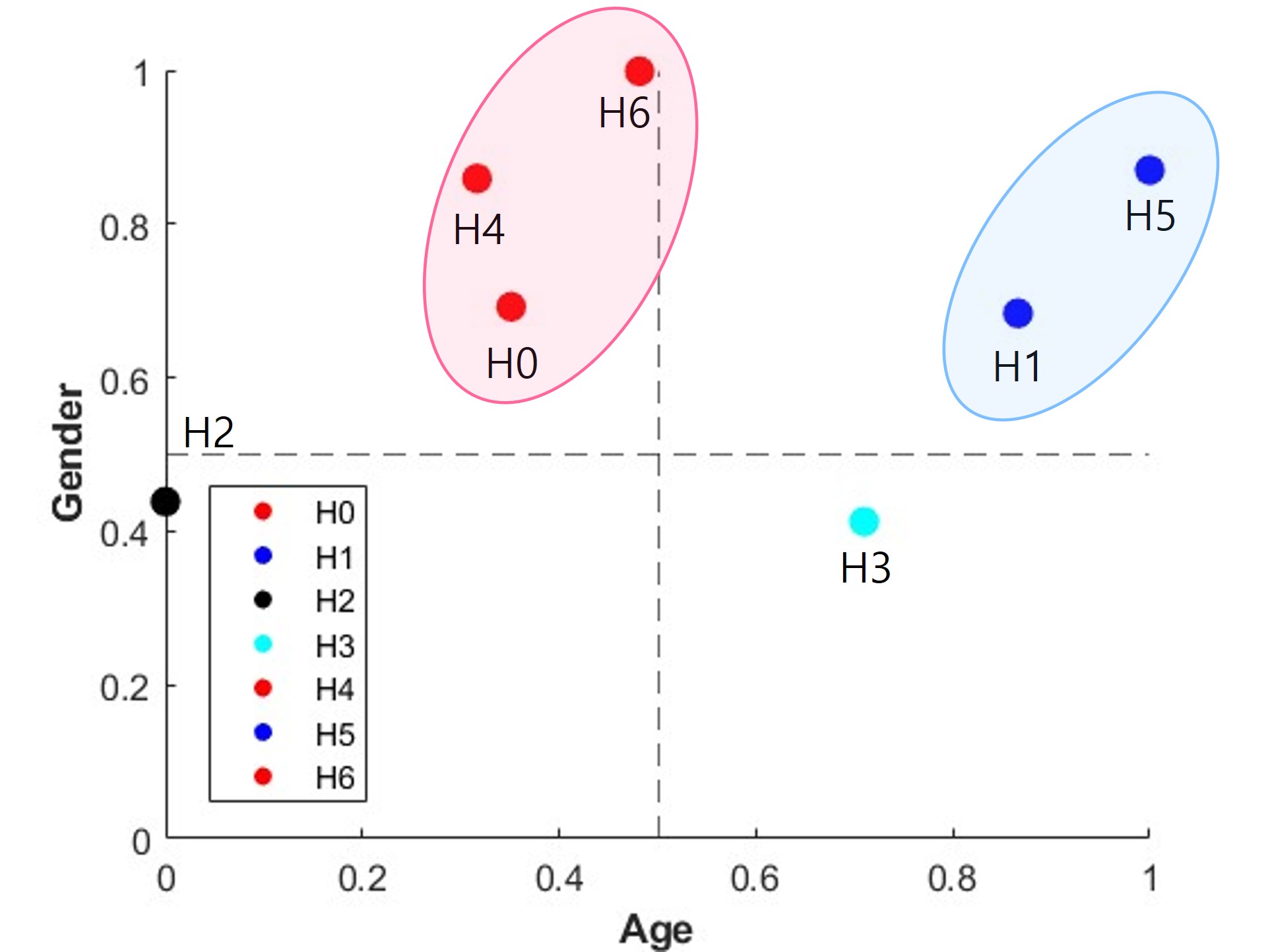}
\label{fig:c_2}
}
\vspace{0.1cm}
\subfigure[]{
\includegraphics[width=0.97\columnwidth]{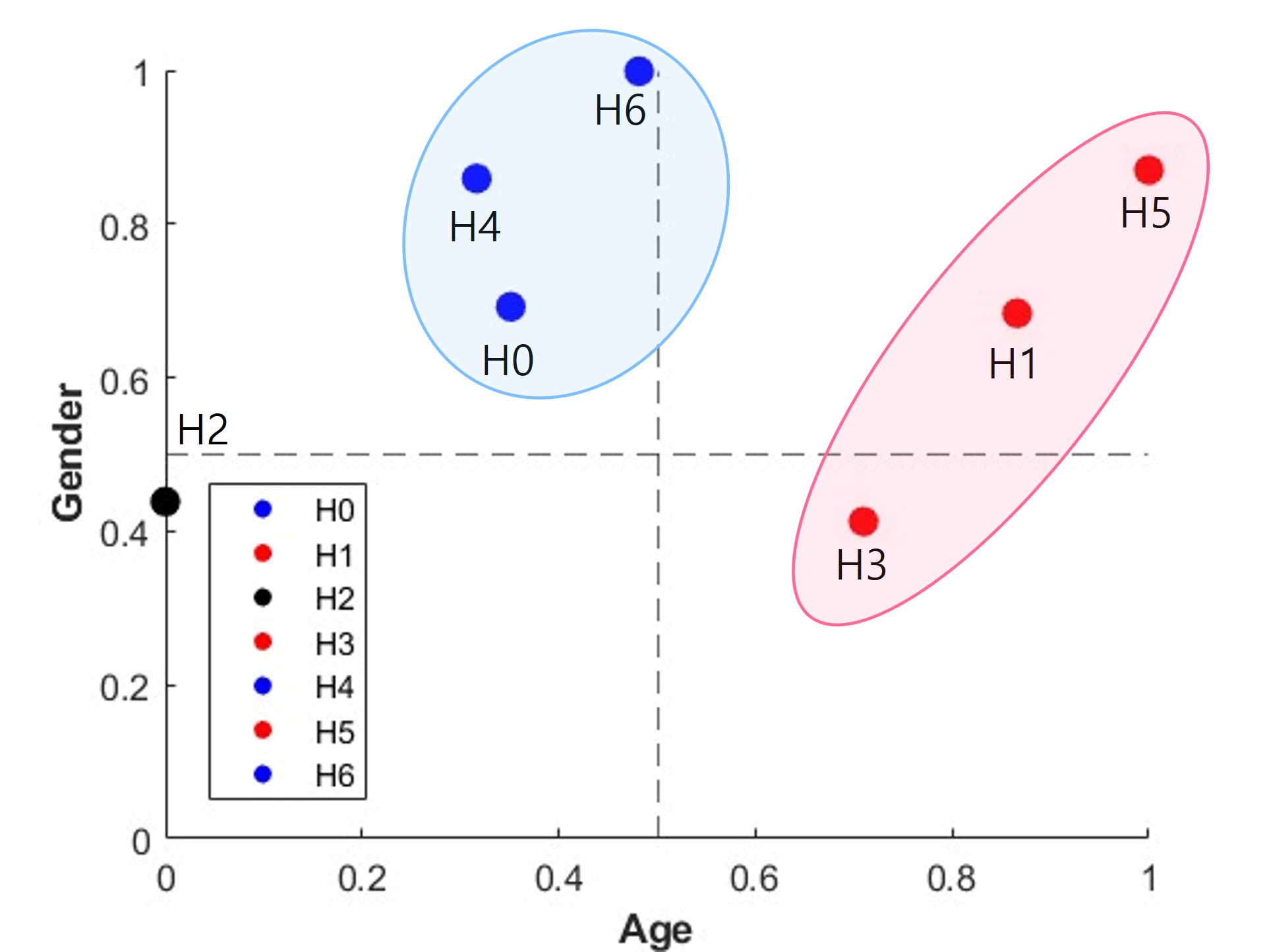}
\label{fig:e_2}
}
\caption{
Clustering Result in Scenario 2: (a) EASB Cluster Algorithm, (b) Cosine Similarity, and (c) Euclidean distance.
}
\label{fig:S2}
\end{figure}

\begin{figure*}[!t]
\centering
\subfigure[]{
\includegraphics[width=\columnwidth]{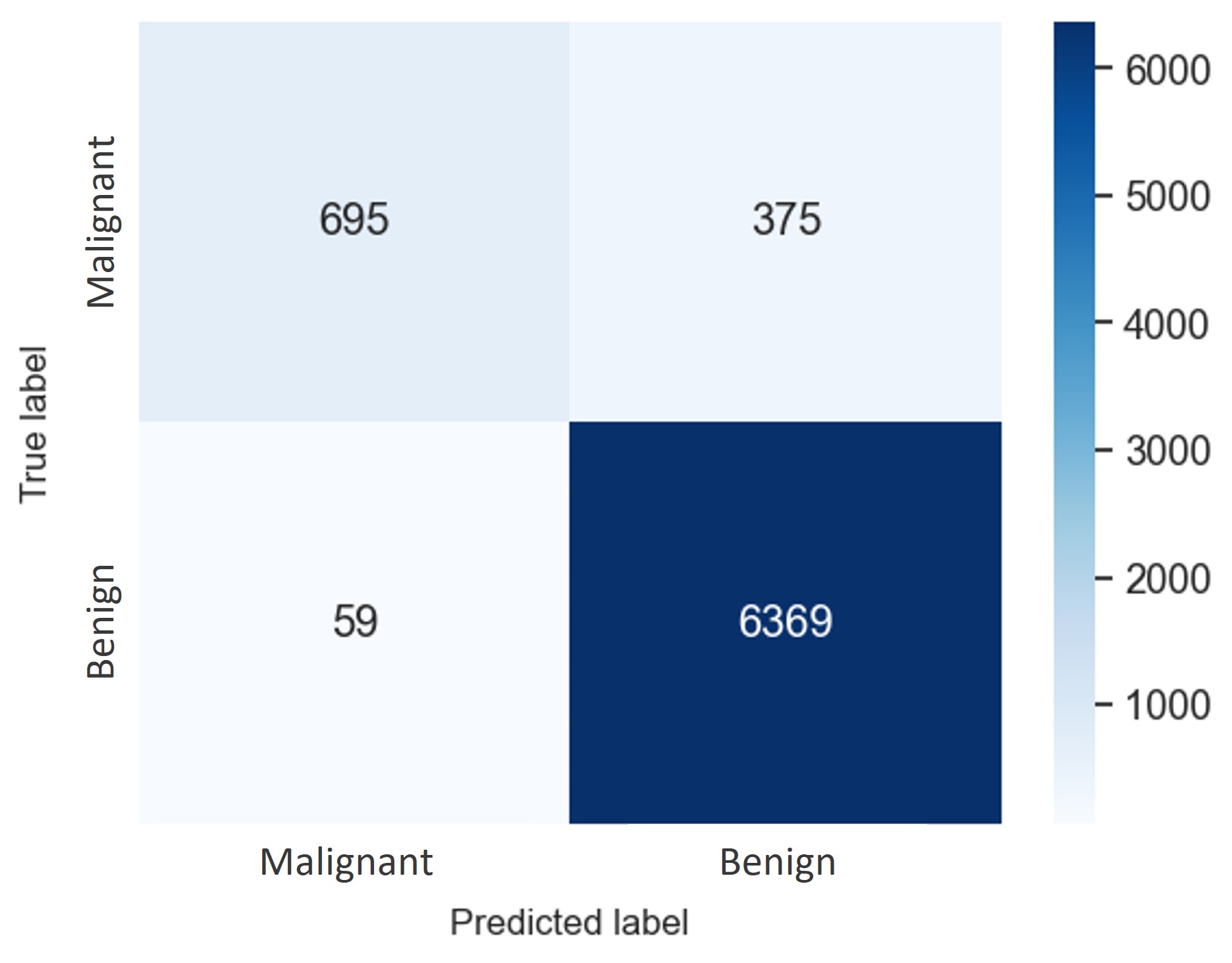}
}
\subfigure[]{
\includegraphics[width=\columnwidth]{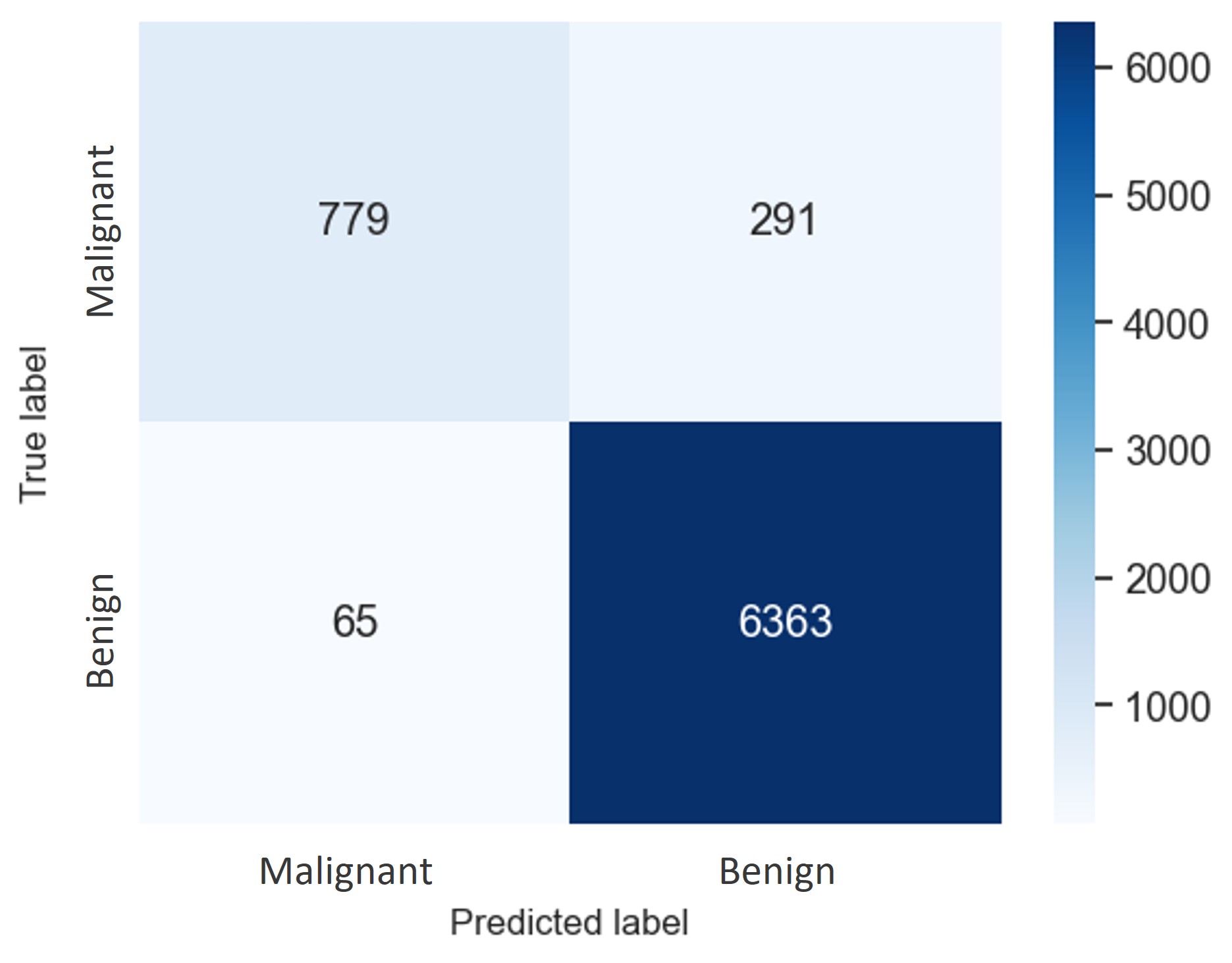}
}
\caption{Details of Confusion Matrix in Scenario 1:
(a) H4 and (b) EASB Cluster.}
\label{fig:c1m}
\end{figure*}

\begin{figure*}[!t]
\centering
\subfigure[]{
\includegraphics[width=\columnwidth]{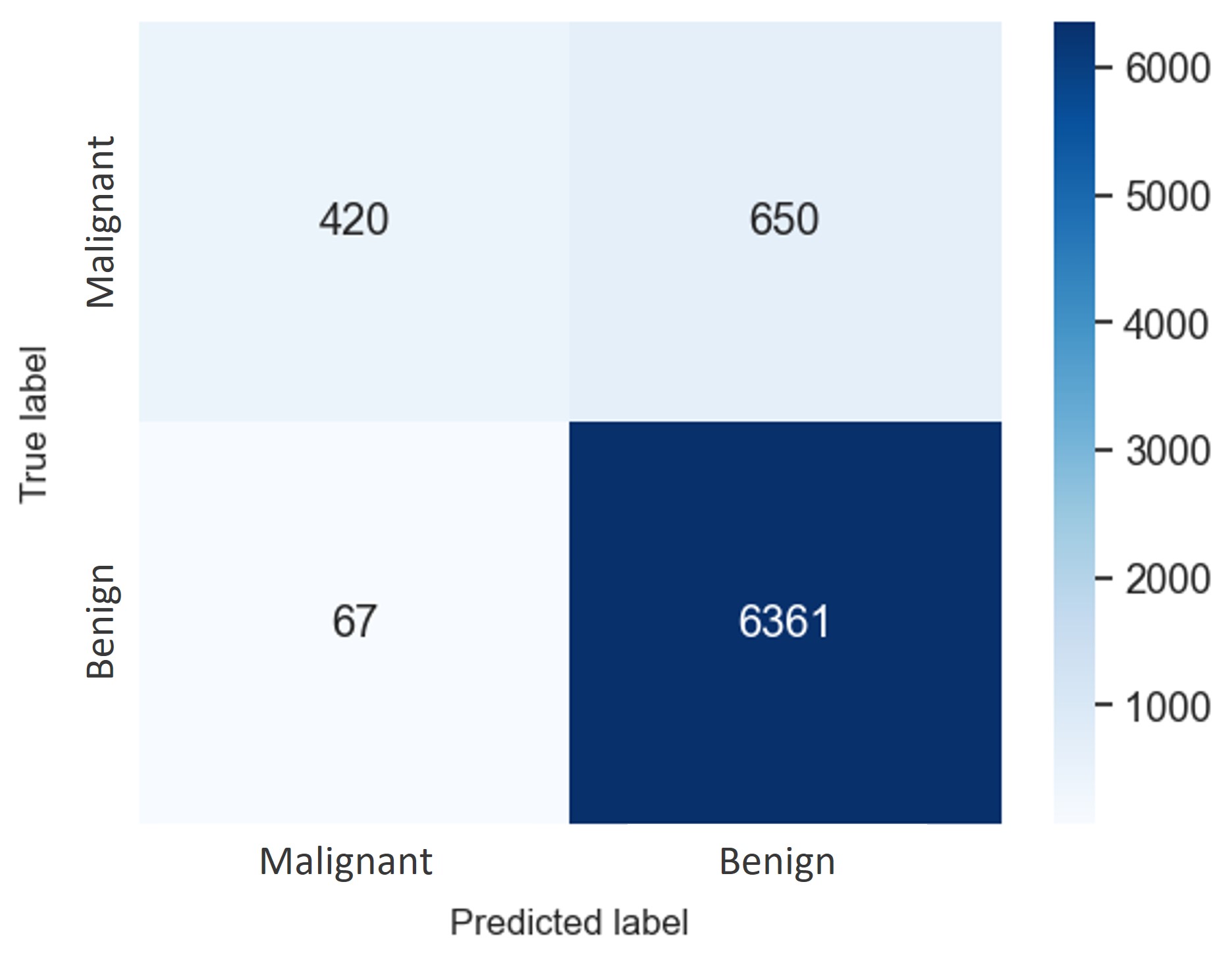}
\label{fig:nrGroup}
}
\subfigure[]{
\includegraphics[width=\columnwidth]{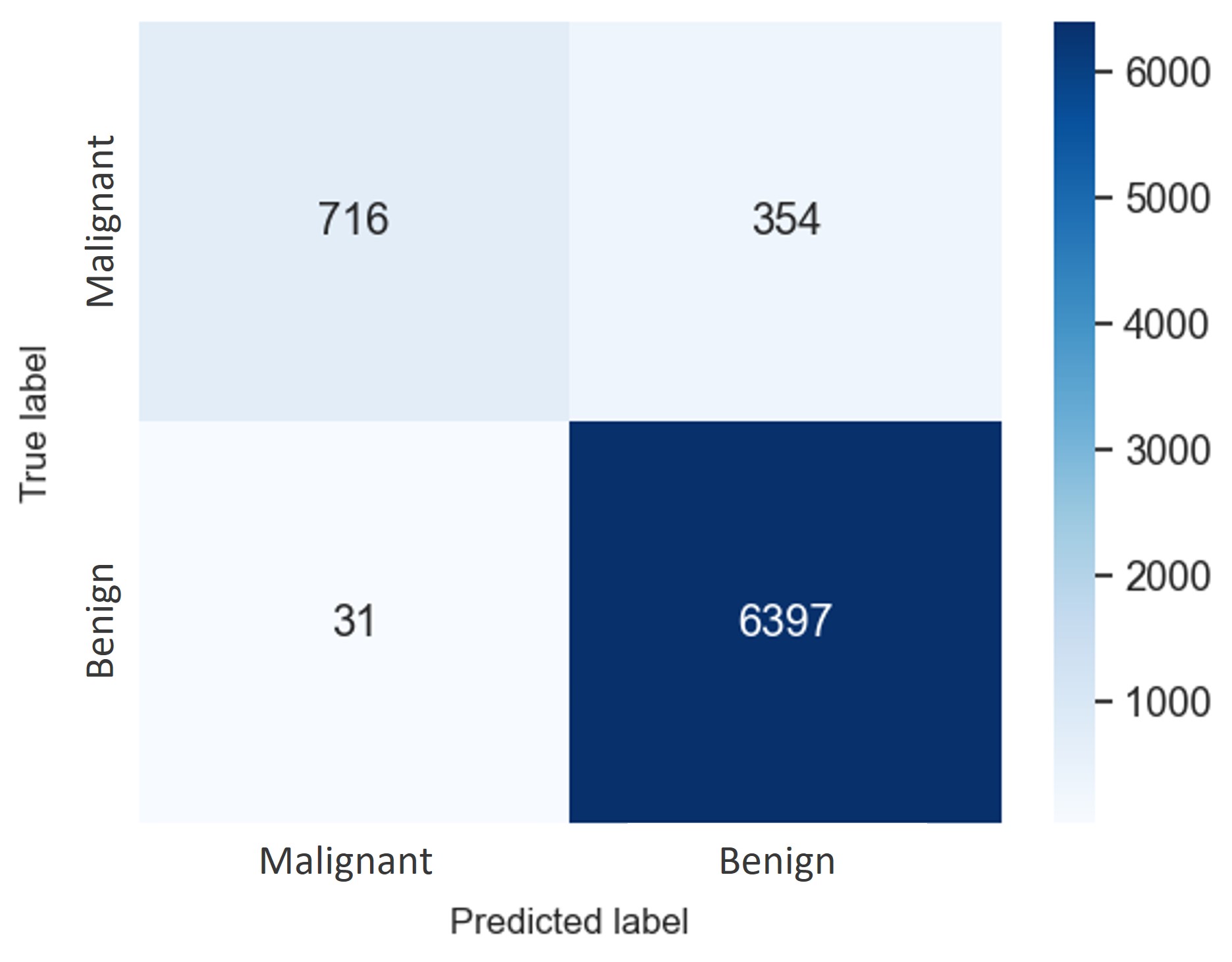}
\label{fig:overallResult}
}
\caption{Confusion Matrix in Scenario 2:
(a) H5 and (b) EASB Cluster.
}
\label{fig:S2m}
\end{figure*}

\subsection{Scenario 2}

\subsubsection{Comparison of Balance}

In scenario 2, the results of clustering based on cosine similarity and Euclidean distance are presented in Fig.~\ref{fig:S2}(b) and (c), respectively. The existing method's clustering based on angle and distance is not balanced clustering by crossing unbalanced hospitals, as demonstrated by both cosine similarity and Euclidean distance making age-unbalanced groups, which are H0, H4, and H6, into one cluster. However, using our algorithm, the age-unbalanced groups (i.e., H0, H4, and H6), and the gender-unbalanced groups (i.e., H3), are combined into one cluster, except for balanced hospitals (i.e., H1 and H5) to address the imbalance. As a result, we are able to create three balanced clusters: H5, H1, and a cluster consisting of H0, H3, H4, and H6, as shown in Fig.~\ref{fig:S2}(a).

Table~\ref{tab:SC2_Comparison} provides the average degree of balance achieved by each method. As shown in Table~\ref{tab:SC2_Comparison}, our proposed algorithm for creating clusters demonstrates superior balance compared to other existing methods, indicating that our approach results in better balance when forming clusters. Therefore, our algorithm performs better in terms of balanced clustering than the existing methods.

\begin{table}[t!]
\centering
\caption{Comparison of Balance in Scenario 2}
\small{
\label{tab:SC2_Comparison}
\resizebox{0.9\columnwidth}{!}{%
\begin{tabular}{c|c|c}
\hline
\textbf{Avg\_Cluster} & \textbf{$H_{Age}[Y]_i$ (\%)} & \textbf{$H_{Gender}[Y]_i$ (\%)} \\ \hline \hline
Cosine             & 55.14 & 64.23 \\ \hline
Euclidean distance & 48.73 & 68.83 \\ \hline
EASB        & 63.29 & 74.67 \\ \hline
\end{tabular}%
}}
\end{table}

\begin{table}[t!]
\centering
\caption{Comparison of Melanoma Detection in Scenario 2}
\label{tab:Scenario 2_result}
\resizebox{\columnwidth}{!}{%
\begin{tabular}{c|c|cccc}
\hline
\multirow{2}{*}{\textbf{\begin{tabular}[c]{@{}c@{}}\rule{0in}{3.2ex} Hospital\\Cluster \end{tabular}}} &
  \multirow{2}{*}{\textbf{\begin{tabular}[c]{@{}c@{}}\rule{0in}{3.2ex} AUROC\\ \normalfont(\%)\end{tabular}}} &
  \multicolumn{4}{c}{\textbf{Confusion Matrix}} \\ \cline{3-6} 
 &
   &
  \multicolumn{1}{c|}{\textbf{\begin{tabular}[c]{@{}c@{}}Accuracy\\ (\%)\end{tabular}}} &
  \multicolumn{1}{c|}{\textbf{\begin{tabular}[c]{@{}c@{}}Precision\\ (\%)\end{tabular}}} &
  \multicolumn{1}{c|}{\textbf{\begin{tabular}[c]{@{}c@{}}Recall\\ (\%)\end{tabular}}} &
  \textbf{\begin{tabular}[c]{@{}c@{}}F1-score\\ (\%)\end{tabular}} \\ \hline \hline
H5 &
  90.60 &
  \multicolumn{1}{c|}{90.44} & 
  \multicolumn{1}{c|}{86.24} &
  \multicolumn{1}{c|}{39.25} &
  53.95 \\ \hline
EASB &
   97.15&
  \multicolumn{1}{c|}{94.87} &
  \multicolumn{1}{c|}{95.85} &
  \multicolumn{1}{c|}{66.92} &
  78.81
   \\ \hline
\end{tabular}%
}
\end{table}

\subsubsection{Comparison of Melanoma Detection}
Our study aims to evaluate the effectiveness of our algorithm for detecting melanoma by comparing its performance on a well-balanced hospital cluster, which contains H0, H3, H4, and H6, and the most balanced hospital cluster (i.e., H5). If the cluster we created has similar or superior performance to H5, the most balanced hospital, the cluster would have been well balanced. The experiment results are shown in Table~\ref{tab:Scenario 2_result}. As shown in Table~\ref{tab:Scenario 2_result}, it can be seen that the cluster we created performs better than H5. Therefore, we can confirm that the balance is well matched with the new similarity formula we created and that melanoma detection is also well performed. In the case of AUROC, the cluster was 97.15\%, superior to 90.65\%, which is for H5. In the case of the Confusion matrix, it was confirmed that the cluster performed better than H5. Specifically, in the case of accuracy, the cluster was 94.87\%, which showed better performance than 90.44\%, which is for H5. In the case of precision, it can be seen that the cluster is 95.85\%, and H5 shows better performance than 86.24\%. In the case of recall, the cluster was 66.92\%, which was superior to 64.95\%, which is for H5. Finally, in F1-score, it can be seen that the cluster is 78.81\%, which is superior to 53.95\%, which is for H5.

Moreover, our research analyzes the performance of H5 and our newly constructed balanced cluster in melanoma detection by examining the confusion matrix in details, as depicted in Fig.~\ref{fig:S2m}(a) and (b). Fig.~\ref{fig:S2m}(a) illustrates that H5 accurately identifies 420 malignant out of 1,070 malignant images however misjudged 650 images as benign. It also correctly identifies 6,361 benign images out of 7,498 however misjudged 67 images as malignant. Fig.~\ref{fig:S2m}(b) shows that our newly constructed cluster accurately identifies 716 malignant out of 1,070 malignant images, whereas misjudged 354 as benign. It also correctly identifies 6,397 benign images out of 7,498 however misjudged 31 as malignant. Finally, our results indicate that our new balanced cluster algorithm outperforms in all situations than H5 in the case study with melanoma detection.

\section{Concluding Remarks}\label{sec:conclusion}

This paper introduces the novel entropy-aware similarity for balanced (EASB) clustering algorithm that considers the balance within the group for the case study with melanoma detection. Unlike general clustering algorithms based solely on similarity, our proposed EASB algorithm considers balance and similarity at the same time. We consider a weight factor in our similarity formulation that can be balanced using entropy and also considered angles and distances. In particular, in the melanoma detection applications we target, gender and age critically impact survival rates, therefore we tries to balance them based on two factors. The proposed EASB algorithm has been evaluated and compared to conventional similarity clustering methods, and our EASB algorithm has proven superior in terms of balance. Its superiority is also demonstrated by comparing melanoma detection performance between the clustered and the most balanced hospital groups. The data-intensive performance evaluation results confirm that the proposed EASB algorithm exhibits superior performance while considering data balance in melanoma detection. 
In addition, this can solve the problem of biased information collection based on the data collection in a single hospital. Furthermore, this EASB algorithm also guarantees the patients' privacy of the hospital dataset because hospitals do not need to share information with all the other hospitals.

As future work, data intensive and performance evaluation with various criteria can be conducted. 
In addition, a new similarity formula that considers data size can be doable, and a balancing degree clustering mechanism can be also considerable for various datasets beyond melanoma detection. Finally, we can find more applications that our proposed EASB algorithm can be utilized which can verify the generalized novelity of our work.

\bibliographystyle{IEEEtran}
\bibliography{ref_aimlab}
\begin{IEEEbiography}[{\includegraphics[width=1in,height=1.25in,clip]{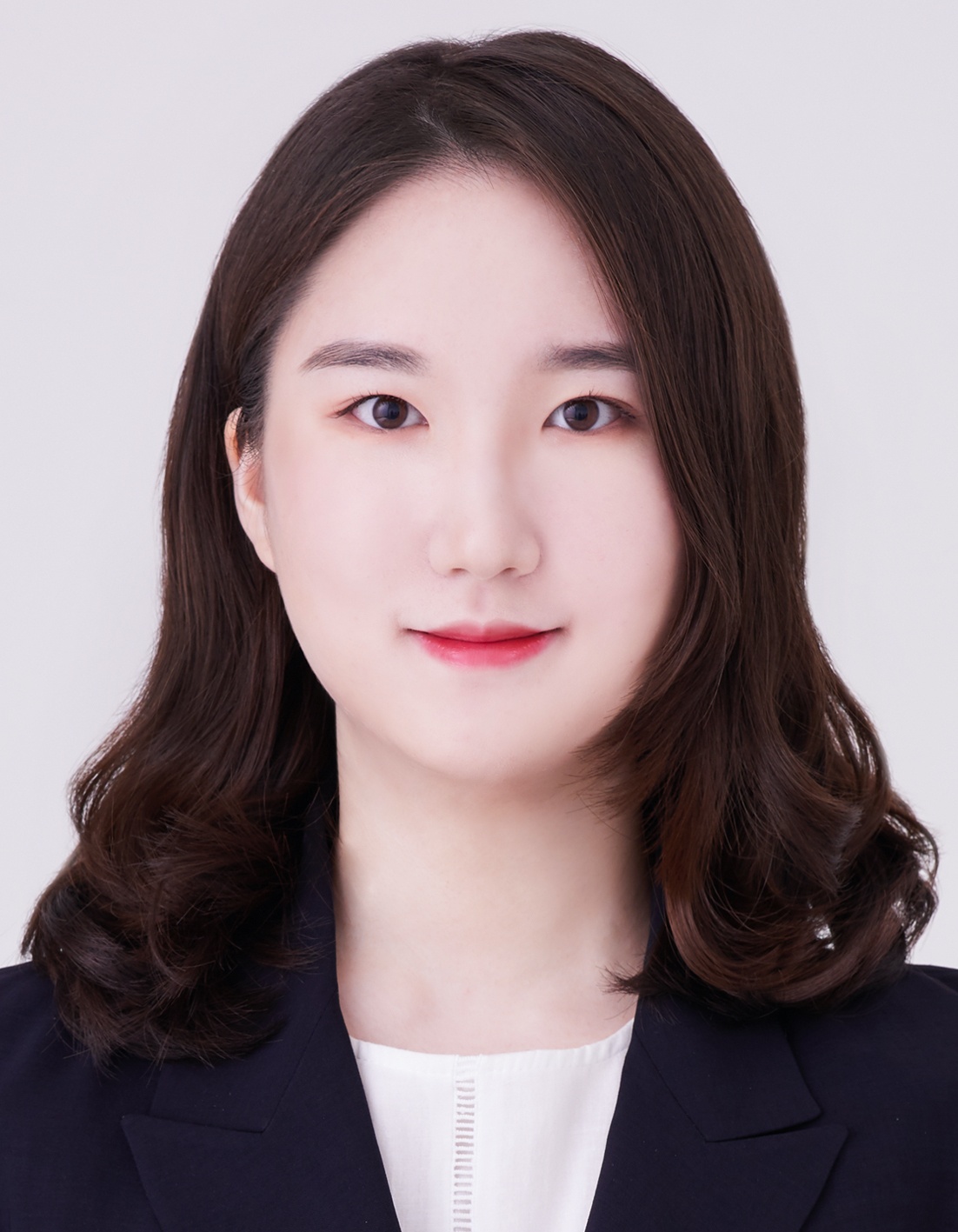}}]{Seok Bin Son} is currently pursuing the M.S. degree in electrical and computer engineering at Korea University, Seoul, Republic of Korea. She received the B.S. degree in information security at Seoul Women's University, Seoul, Republic of Korea. Her research focuses include deep learning algorithms and their applications to information security.
\end{IEEEbiography}

\begin{IEEEbiography}[{\includegraphics[width=1in,height=1.25in,clip]{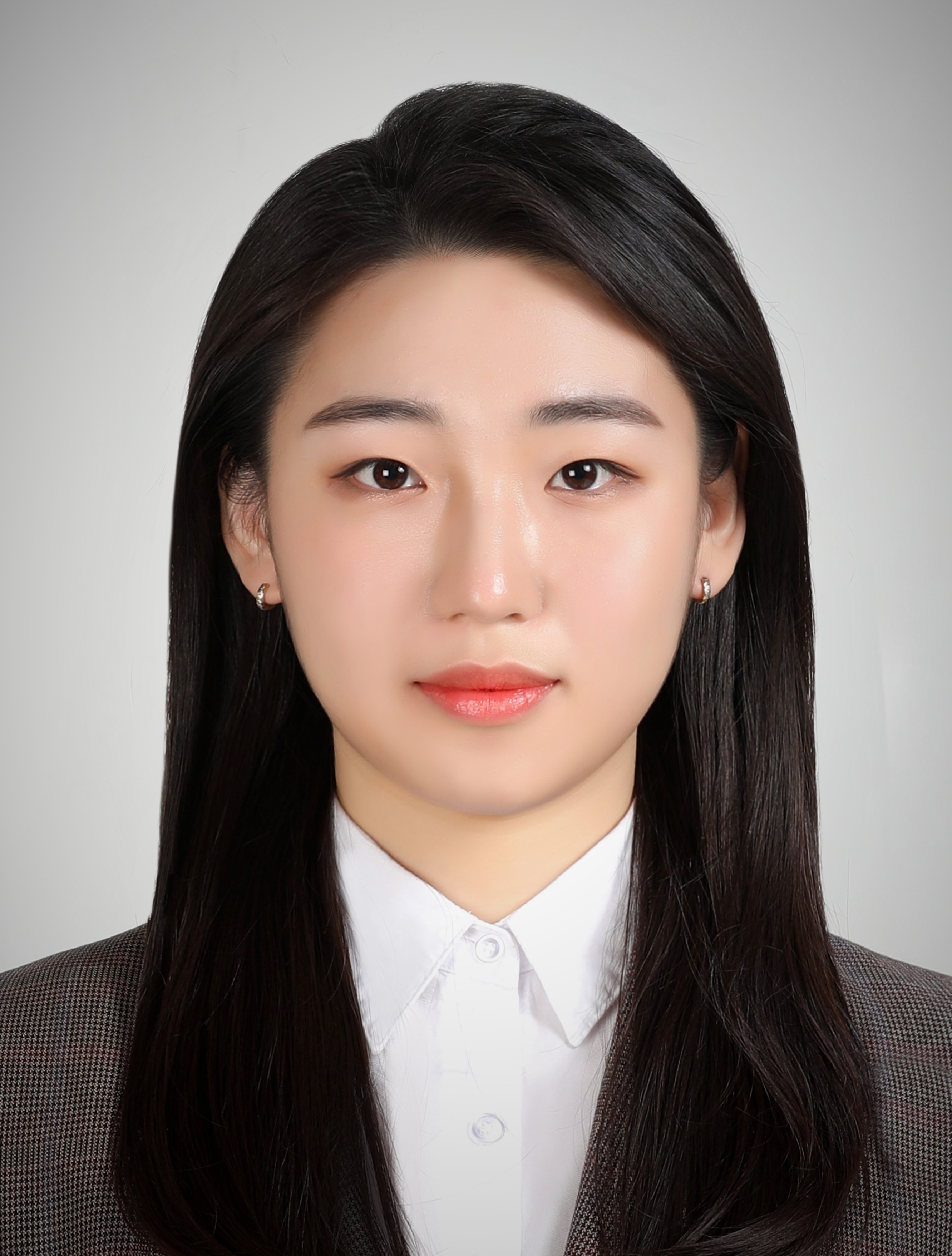}}]{Soohyun Park} 
is currently pursuing the Ph.D. degree in electrical and computer engineering at Korea University, Seoul, Republic of Korea. She received the B.S. degree in computer science and engineering from Chung-Ang University, Seoul, Republic of Korea, in 2019. Her research focuses include deep learning algorithms and their applications to autonomous mobility and connected vehicles. 

She was a recipient of the IEEE Vehicular Technology Society (VTS) Seoul Chapter Award in 2019.
\end{IEEEbiography}

\begin{IEEEbiography}[{\includegraphics[width=1in,height=1.25in,clip,keepaspectratio]{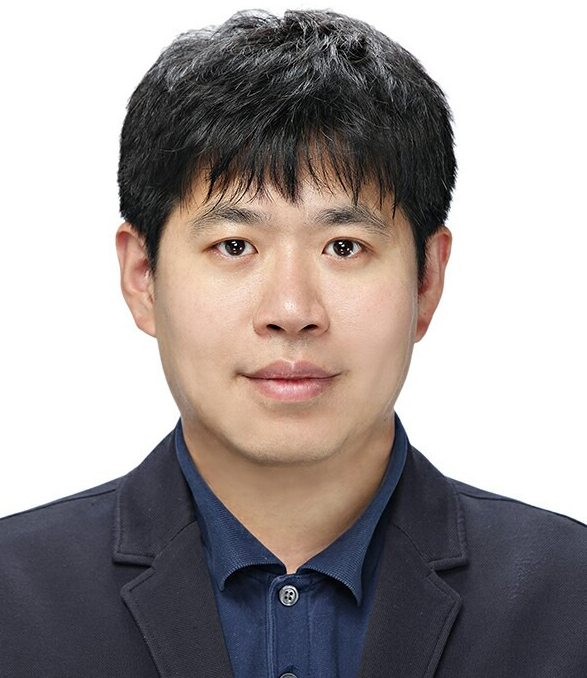}}]{Joongheon Kim}
(M'06--SM'18) has been with Korea University, Seoul, Korea, since 2019, where he is currently an associate professor at the School of Electrical Engineering and also an adjunct professor at the Department of Communications Engineering (established/sponsored by Samsung Electronics) and the Department of Semiconductor Engineering (established/sponsored by SK Hynix). He received the B.S. and M.S. degrees in computer science and engineering from Korea University, Seoul, Korea, in 2004 and 2006; and the Ph.D. degree in computer science from the University of Southern California (USC), Los Angeles, CA, USA, in 2014. Before joining Korea University, he was a research engineer with LG Electronics (Seoul, Korea, 2006--2009), a systems engineer with Intel Corporation (Santa Clara, CA, USA, 2013--2016), and an assistant professor of computer science and engineering with Chung-Ang University (Seoul, Korea, 2016--2019). 

He serves as an editor for \textsc{IEEE Transactions on Vehicular Technology}, \textsc{IEEE Transactions on Machine Learning in Communications and Networking}, and \textsc{IEEE Communications Standards Magazine}. He is also a distinguished lecturer for \textit{IEEE Communications Society (ComSoc)} and \textit{IEEE Systems Council}.

He was a recipient of Annenberg Graduate Fellowship with his Ph.D. admission from USC (2009), Intel Corporation Next Generation and Standards (NGS) Division Recognition Award (2015), \textsc{IEEE Systems Journal} Best Paper Award (2020), IEEE ComSoc Multimedia Communications Technical Committee (MMTC) Outstanding Young Researcher Award (2020), IEEE ComSoc MMTC Best Journal Paper Award (2021), and Best Special Issue Guest Editor Award by \textit{ICT Express (Elsevier)} (2022). He also received several awards from IEEE conferences including IEEE ICOIN Best Paper Award (2021), IEEE Vehicular Technology Society (VTS) Seoul Chapter Awards (2019, 2021, and 2022), and IEEE ICTC Best Paper Award (2022). 
\end{IEEEbiography}
\EOD
\end{document}